\documentclass[aps,superscriptaddress,floats,showpacs,floatfix,onecolumn]{revtex4}
\usepackage{bm}
\usepackage{graphicx}
\usepackage{subfigure}
\usepackage{amsmath}
\usepackage{bbm}
\usepackage{amssymb}
\usepackage{color}
\usepackage{hyperref}
\usepackage[a4paper,left=1.5cm, right=1.3cm, top=3.0cm,bottom=2cm]{geometry}

\begin{document}
\newcommand{\Nu}{{\rm Nu}}   
\newcommand{\Rey}{{\rm Re}}   
\newcommand{\Ra}{{\rm Ra}}   
\newcommand{\Pra}{{\rm Pr}}   
\newcommand{\ur}{u_{\rm rms}}

\title{Lagrangian heat transport in turbulent three-dimensional convection}
\author{Philipp P. Vieweg}
\affiliation{Institut f\"ur Thermo- und Fluiddynamik, Technische Universit\"at Ilmenau, Postfach 100565, D-98684 Ilmenau, Germany}
\author{Christiane Schneide}
\affiliation{Institut f\"ur Mathematik und ihre Didaktik, Leuphana Universit\"at L\"uneburg, D-21335 L\"uneburg, Germany}
\author{Kathrin Padberg-Gehle}
\affiliation{Institut f\"ur Mathematik und ihre Didaktik, Leuphana Universit\"at L\"uneburg, D-21335 L\"uneburg, Germany}
\author{J\"org Schumacher}
\affiliation{Institut f\"ur Thermo- und Fluiddynamik, Technische Universit\"at Ilmenau, Postfach 100565, D-98684 Ilmenau, Germany}
\date{\today}

\begin{abstract}
Spatial regions that do not mix effectively with their surroundings and thus contribute less to the heat transport in fully turbulent three-dimensional Rayleigh-B\'{e}nard flows are identified by Lagrangian trajectories that stay together for a longer time. These trajectories probe Lagrangian coherent sets (CS) which we investigate here in direct numerical simulations in convection cells with square cross section of aspect ratio $\Gamma = 16$, Rayleigh number $\Ra = 10^{5}$, and Prandtl numbers $\Pra = 0.1, 0.7$ and $7$. The analysis is based on $N=524,288$ Lagrangian tracer particles which are advected in the time-dependent flow. Clusters of trajectories are identified by a graph Laplacian with a diffusion kernel, which quantifies the connectivity of trajectory segments, and a subsequent sparse eigenbasis approximation (SEBA) for cluster detection.  The combination of graph Laplacian and SEBA leads to a significantly improved cluster identification that is compared with the large-scale patterns in the Eulerian frame of reference. We show that the detected CS contribute by a third less to the global turbulent heat transport for all investigated $\Pra$ compared to the trajectories in the spatial complement. This is realized by monitoring Nusselt numbers along the tracer trajectory ensembles, a dimensionless {\em local} measure of heat transfer.
\end{abstract}
\maketitle

\section{Introduction} 
The investigation of transport and mixing properties in complex dynamical systems in the Lagrangian frame of reference has received an increasing interest in the past two decades and many different identification methods have been developed and applied to fluid flows, see \cite{Allshouse2015,Haller2015,Hadjighasem2017} for recent reviews. Central to these approaches is the concept of a Lagrangian {\em coherent set} (CS) \cite{FrLlSa10,Froyland2013,Allshouse2015,Karrasch2017}, representing a region in the fluid volume that only weakly mixes with its surrounding and which often stays regularly shaped (non-filamented) under the evolution of the flow. Coherent sets were originally introduced on the basis of transfer operators \cite{FrLlSa10,Froyland2013}, but in the past few years several approaches were proposed that use spatio-temporal clustering algorithms applied to Lagrangian trajectory data \cite{Froyland_Padberg_2015,Hadjighasem2016,Banisch2017,Schlueter2017,Padberg2017,FroylandJunge2018,Schneide2018}. These algorithms aim at identifying coherent sets as groups of tracer trajectories that remain close to each other or behave in a similar manner in the time interval under investigation. In turbulent convection flows, the heat transport from the bottom to the top across an extended layer or a closed vessel is one of the essential processes that require a deeper understanding \cite{Ahlers2009,Chilla2012} in view to the numerous geophysical \cite{Stevens2005}, astrophysical \cite{Schumacher2020} and technological \cite{Kelley2018} applications. One first step is to identify the spatial sets that contribute least to this transport and to relate them to the large-scale structures which are observed in the Eulerian frame of reference.

\begin{figure}
\begin{center}
\includegraphics[scale=1.0]{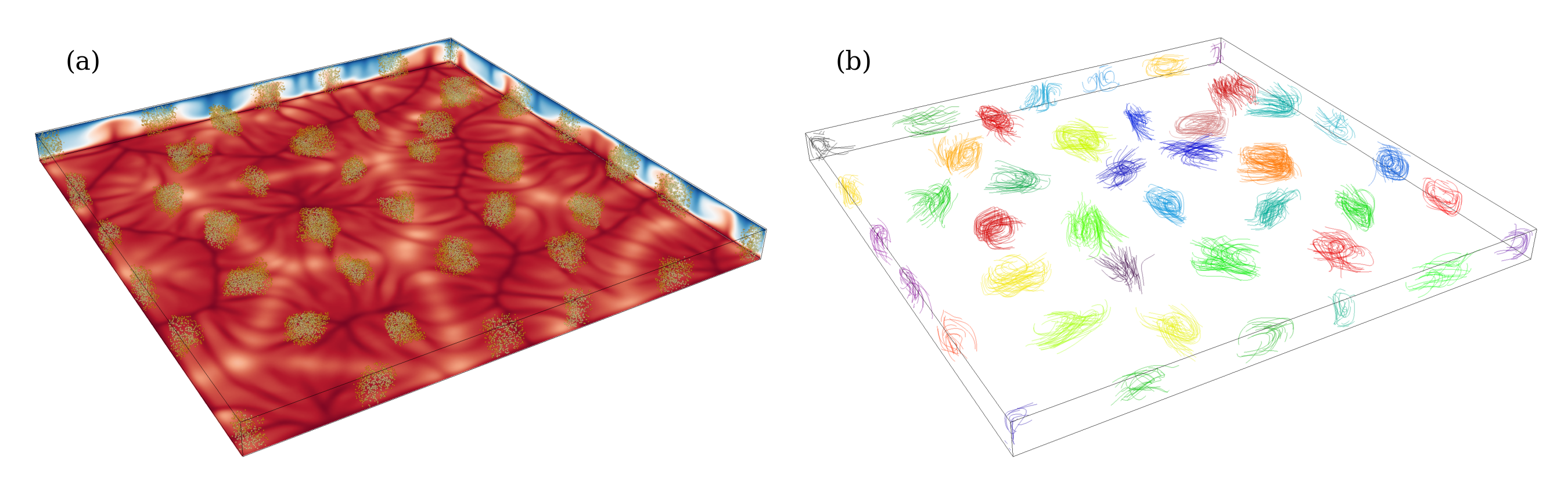}
\caption{Visualization of Lagrangian coherent sets. (a) The $40,369$ Lagrangian tracers which correspond to Lagrangian CS are shown for $\Ra = 10^{5}$, $\Pra = 7$, and $\Gamma = 16$ (Run 3). The highlighted particles comprise the cluster cores during the chosen time interval $[t_0, t_0+\Delta t]$ with $\Delta t=17$, while the shown tracer positions correspond to the time $t_0+\Delta t/2$. Contours of the temperature field, which is averaged over the same time interval, are displayed at three faces. The bottom plane stands for $\langle T(z = 0.05) \rangle_{\Delta t}$. (b) A representative subset of $1,600$ Lagrangian trajectories is shown. They form the different clusters in (a).}
\label{fig:3d_clusters_RBC}
\end{center}
\end{figure}

In the present work, we identify CS as spatial regions of reduced turbulent heat transfer in three-dimensional Rayleigh-B\'{e}nard convection (RBC) using a set-based approach. Our Lagrangian analysis of turbulent RBC starts with a graph Laplacian that originates, as in \cite{Hadjighasem2016}, from the time-averaged distances between particles and uses a Gaussian kernel in the spectral clustering approach (the latter of which can also be interpreted as a random walk or diffusion process on the data \cite{Meila2001,Coifman2005,Coifman2006}). Three different Prandtl numbers ${\rm Pr}$, a dimensionless parameter that relates the viscosity of the fluid to its temperature diffusivity, are considered. To extract Lagrangian coherent sets from spectral properties of the graph Laplacian, we apply the recently developed sparse eigenbasis approximation (SEBA) \cite{Froyland2019}. The combination of these methods allows us to disentangle the contribution of the tracer trajectories that are trapped in CS to the overall heat transfer in comparison to the rest, thus extending our recent Lagrangian studies of RBC  \cite{Schneide2018,Schneide2019,Kluenker2020}. Figure \ref{fig:3d_clusters_RBC} illustrates these coherent sets and corresponding representative trajectory segments. It turns out that these regions contribute least to the turbulent heat transfer. We will show this by monitoring the local Nusselt number along individual Lagrangian trajectory segments.  

The three-dimensional Boussinesq equations are solved by the spectral element method nek5000 \cite{Fischer1997,Scheel2013,Schneide2018}. Details of all simulations are summarized in Table I. The equations are made dimensionless by the layer height $H$, wall-to-wall temperature difference $\Delta T$, and free-fall velocity $U_{f} = \sqrt{g \alpha \Delta T H}$ with acceleration due to gravity $g$ and isobaric expansion coefficent $\alpha$,
\begin{align}
\label{eq:continuity_equation}
\nabla \cdot {\bm u} &= 0\,,\\
\label{eq:Navier-Stokes_equation}
\frac{\partial {\bm u}}{\partial t} + \left( {\bm u} \cdot \nabla \right) {\bm u} &= - \nabla p + \sqrt{\frac{\Pra}{\Ra}} \nabla^{2} {\bm u} + T {\bm e_{z}} \,,\\
\label{eq:energy_equation}
\frac{\partial T}{\partial t} + \left( {\bm u} \cdot \nabla \right) T &= \frac{1}{\sqrt{\Ra \Pra}} \nabla^{2} T\,.
\end{align}
We consider a closed cell with square horizontal cross section $L_{\rm hor}\times L_{\rm hor}$ and an aspect ratio of $\Gamma=L_{\rm hor}/H=16$. No-slip boundary conditions for the velocity field are applied at all walls. The sidewalls are thermally insulated and $T_{\rm bottom}=1$ and $T_{\rm top}=0$ are chosen.  The volume is covered by more than 440,000 spectral elements. The Rayleigh number is given by ${\rm Ra} = g \alpha \Delta T H^{3}/(\nu \kappa)$ and the Prandtl number ${\rm Pr}=\nu/\kappa$. Here, $\nu$ is the kinematic viscosity and $\kappa$ is the thermal diffusivity. We advect $N=524,288$ massless Lagrangian tracer particles by $\dot{\bm X}_i={\bm u}({\bm X}_i,t)$ for $i=1,\dots,N$ together with the time-dependent flow. The cluster analysis starts when the Lagrangian tracer particles are uniformly distributed across the cell after the initial seeding.

It is well-known that turbulent convection flows in extended domains get organized into prominent large-scale patterns which are termed turbulent superstructures of convection, see e.g. \cite{Stevens2018,Pandey2018,Green2020,Krug2020,Vieweg2020}. Table I compares the characteristic lengths and times for the Eulerian ($E$) and Lagrangian ($L$) frames (for comparison see also refs. \cite{Pandey2018,Schneide2018}). The Eulerian characteristic length $\bar{\lambda}^{E}$ is the wavelength that corresponds to the wavenumber $k^{\ast}$ at which the time-averaged Fourier spectrum of the vertical velocity component in the midplane becomes maximal, whereas the Eulerian characteristic time is a mean turnover time calculated by $\bar{\tau}^{E} \approx \pi (\lambda^{E} + 2)/(4\ur)$. The corresponding Lagrangian values are obtained as means of the probability density functions (PDFs) which are taken over the whole tracer ensemble. Therefore, we determine $\lambda^L$ as four times the horizontal travel distance of each tracer between two successive intersections of the midplane. Thus it probes on average the typical counterrotating double roll structure. In comparison, $\tau^L$ is the time each particle needs to complete a full turnover, which is probed by passing the heights $z = 0.2$ and $0.8$. Despite of the extended tails of all distributions, the mean values $\bar{\lambda}^{L}$ and $\bar{\tau}^{L}$ are found to be close to the corresponding characteristic scales of the Eulerian frame, confirming the consistency of our analyses. 

\begin{table*}[t]
\label{tab:Eulerian_Lagrangian_scales}
\renewcommand{\arraystretch}{1.5}
\centering
\begin{tabular}{cccccccccccccccc}
\hline\hline
Run  & \multicolumn{1}{c}{$\Ra$}	& \multicolumn{1}{c}{$\Pra$}	& \multicolumn{1}{c}{$\Nu^{E}$}	& \multicolumn{1}{c}{$\bar{\tau}^{E}$} 	& \multicolumn{1}{c}{$\bar{\lambda}^{E}$} 	& \multicolumn{1}{c}{$\Delta t^{L}$} 	& \multicolumn{1}{c}{$\bar{\tau}^{L}$} 	& \multicolumn{1}{c}{$\bar{\lambda}^{L}$} & \multicolumn{1}{c}{$\epsilon$} & \multicolumn{1}{c}{$\Delta t$} 	& \multicolumn{1}{c}{$N_{\rm CS}$} 	& \multicolumn{1}{c}{$\overline{\Nu}^{L}_{\rm CS}$} 	& \multicolumn{1}{c}{$\overline{\Nu}^{L}_{\rm RP}$} 	& \multicolumn{1}{c}{$\xi_{p}$} 	& \multicolumn{1}{c}{$\xi_{q}$}\\
\hline
1 & $10^5$ & $0.1$ 	& $3.50$ 	& $10.6$ 	& $3.4$ 	& $133$ 	& $13.7 \pm 8.7$ 	& $3.7 \pm 2.1$ & $49/800$ 	& $3.25$ 	& $75$ 	& $2.89$ 	& $3.54$ 	& $7.9 \%$ 	& $6.6 \%$ \\
2 & $10^5$ & $0.7$ 	& $4.13$ 	& $18.3$ 	& $3.6$ 	& $234$ 	& $21.7 \pm 14.5$ 	& $3.6 \pm 2.1$ & $9/200$ 	& $5.50$ 	& $80$ 	& $2.71$ 	& $4.41$ 	& $8.7 \%$ 	& $5.5 \%$ \\
3 & $10^5$ & $7.0$ 	& $4.18$ 	& $63.4$ 	& $5.1$ 	& $700$ 	& $68.6 \pm 51.1$ 	& $5.0 \pm 2.4$ & $9/200$ 	& $17.00$ 	& $40$ 	& $2.87$ 	& $4.30$ 	& $7.6 \%$ 	& $5.2 \%$ \\
\hline
\hline
\end{tabular}
\caption{Parameters and global statistical measures of the simulations. The Rayleigh number $\Ra$, the Prandtl number $\Pra$, the global Nusselt number $\Nu^{E}$, the characteristic turnover time $\bar{\tau}^{E}$ and the characteristic length $\bar{\lambda}^{E}$ in the Eulerian (superscript $E$) frame of reference,  the total time of Lagrangian (superscript $L$) analysis $\Delta t^{L}$, the mean Lagrangian turnover time $\bar{\tau}^{L}$, and the characteristic length $\bar{\lambda}^{L}$ are listed. Error bars follow from standard deviation. Furthermore, we provide the diffusion kernel scale $\epsilon$, the width $\Delta t$ of each time window used to compute the time-averaged distance $r_{ij}$, the approximate number of detected Lagrangian coherent sets $N_{\rm CS}$, the average Nusselt numbers related to the transport across the coherent sets, $\Nu^{L}_{CS}$, and the complement, $\Nu^{L}_{\rm RP}$ (with RP=remaining particles), the mean fraction of Lagrangian particles in coherent sets, $\xi_{p}$, and their mean contribution to heat transport, $\xi_{q}$. The threshold value is always $\zeta=0.94$.}
\end{table*}

\section{Spectral analysis of graph Laplacian and SEBA clustering}
Material transport is subject to turbulent dispersion which will destroy Lagrangian coherence if the observation time is sufficiently long.  We aim to identify subsets of our $N$ Lagrangian trajectories that stay close together for a longer transient period $\Delta t$. To this end, we apply  a spectral clustering approach to the discrete data and explore their connectivities. Specifically, we use the time-averaged distance proposed in \cite{Hadjighasem2016}
\begin{equation}
\label{distance}
r_{ij} = \langle \left| {\bm X}_{i}(t) - {\bm X}_{j}(t) \right| \rangle_{\Delta t}
\end{equation}
between mutual Lagrangian trajectories ${\bm X}_i(t)$ and ${\bm X}_j(t)$ with the time average $\langle \cdot \rangle_{\Delta t}$ taken from $t_0$ to $t_0+\Delta t$ and obtain the graph Laplacian $N\times N$ matrix 
\begin{equation}
\label{eq:def_graph_Laplacian}
{\bm L} = \frac{1}{\epsilon} ({\bm P}- \pmb{\mathbb{I}})\,,
\end{equation}
where $\epsilon$ is the kernel scale, ${\bm P}$ the diffusion matrix and $\pmb{\mathbb{I}}$ the identity matrix. In a nutshell, the entry $P_{ij}$ encodes the probability of switching in a Markov chain from state (Lagrangian trajectory) $i$ to $j$. Thus, the matrix ${\bm L}$ generates a random walk on our trajectory data points. In more detail, the entries of ${\bm P}$ are 
\begin{equation}
\label{eq:def_diffusion_matrix}
P_{ij} = \dfrac{\hat{K}_{ij}}{\sum_{j=1}^N \hat{K}_{ij}} \quad\text{with}\quad 
 \hat{K}_{ij}=\dfrac{K_{ij}}{k_{\epsilon,i}k_{\epsilon,j}} \quad\mbox{and}\quad
K_{ij} =
\begin{cases}
	\exp \left(- r_{ij}^2/\epsilon \right), 	& r_{ij} \leq \delta \\
	0,						        			& r_{ij} > \delta
\end{cases}\,.
\end{equation}
Here, $\delta = \sqrt{2 \epsilon}$ is the cut-off, $k_{\epsilon, i}=\sum_{k=1}^N K_{ik}$ represents the pre-normalization, and ${\bm K}$ is a diffusion kernel matrix. Even though we work with a large number of tracers, which implies that ${\bm K}$ has $N^2=(2 \times 512^{2})^{2} \sim 10^{11}$ entries, the cut-off in \eqref{eq:def_diffusion_matrix} leads to sparsities above $99 \%$ with no loss of accuracy. This allows to exploit very efficient k-d tree data structures \footnote{This is done by usage of the SciPy class scipy.spatial.cKDTree. The documentation can be found at https://docs.scipy.org/doc/scipy/reference/generated/scipy.spatial.cKDTree.html} by computing first the instantaneous Euclidean distances using an intermediate threshold $\delta_{int} > \delta$, then averaging these pairwise distances over several time steps and applying eventually the final cut-off $\delta$ to eliminate all connections that exceed the time-averaged distance threshold ($r_{ij} > \delta$).
The connectivity of the graph depends on the kernel scale $\epsilon$, the time window $\Delta t$, and the number of Lagrangian trajectories $N$. While $N$ is left unchanged, we found in prior parameter studies that $\Delta t \approx 0.25 \bar{\tau}^{L}$ with a kernel scale $\epsilon$ as small as possible (such that the network still remains connected) gives the best results. A decrease of $\Pra$ generates turbulence with a larger Reynolds number. Thus $\epsilon$ depends also on $\Pra$.
\begin{figure}[t]
\begin{center}
\includegraphics[scale=1.0]{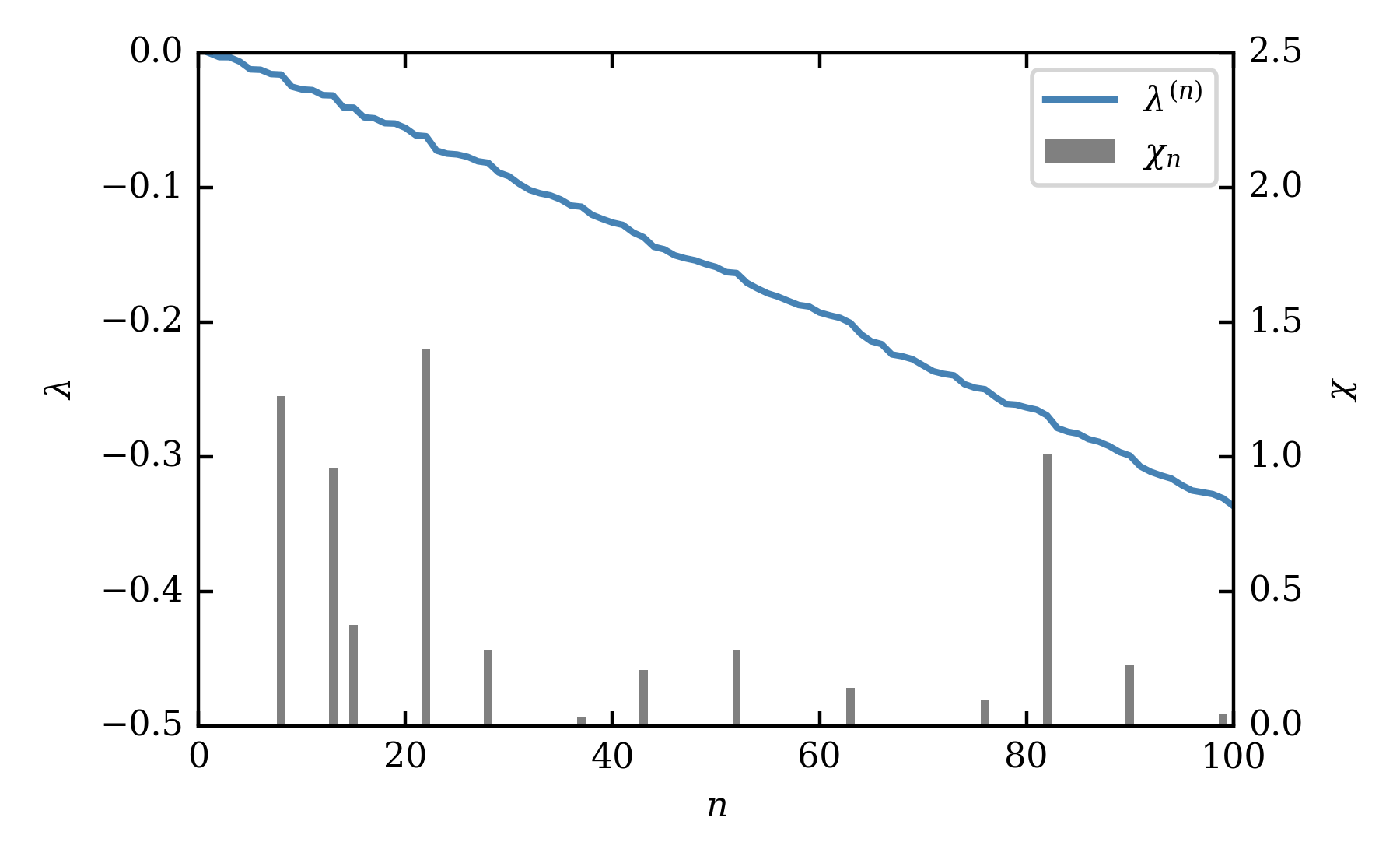}
\caption{A typical eigenspectrum (from top downwards) resulting for the graph Laplacian matrix ${\bm L}$ in Run 2. In order to highlight eigengaps between two subsequent eigenvalues $\lambda^{(n)}$, indicator $\chi_n$ is used, see eq. \eqref{eq:eigengap_detection}.}
\label{fig:eigenspectrum}
\end{center}
\end{figure}
We consider the following eigenvalue problem for the graph Laplacian
\begin{equation}
\label{eq:eigenvalue_problem}
{\bm L} v_n = \lambda^{(n)} v_n\,,
\end{equation}
with the eigenvalues $\lambda^{(n)}$ and corresponding eigenvectors $v_n$. We compute the leading $N_l=250\ll N$ eigenvalues and -vectors for $10$ disjoint time intervals in each of the three runs, where about $50$ equally spaced snapshots are used to evaluate the time-averaged distances $r_{ij}$ in each time window as given in \eqref{distance}. Stronger gaps in the eigenvalue spectrum of ${\bm L}$ indicate potential numbers of clusters as discussed in refs.  \cite{Banisch2017,Froyland2019,Schneide2018,Schneide2019}. We apply the following finite difference ratio to detect if the difference (or gap) from $\lambda^{(n)}$ to $\lambda^{(n+1)}$ is larger than the average difference between the last $k = 4$ predecessive eigenvalues of $\lambda^{(n)}$ and the next $k = 4$ successive eigenvalues of $\lambda^{(n+1)}$, respectively, 
\begin{equation}
\label{eq:eigengap_detection}
\chi_{n}  = \dfrac{8 \left[ \lambda^{(n+1)} - \lambda^{(n)} \right]}{\left[ \lambda^{(n)}-\lambda^{(n-4)} \right] + \left[ \lambda^{(n+5)}-\lambda^{(n+1)} \right]}-1 \quad \mbox{with} \quad 5 \leq n \le N_l-5 \,.
\end{equation}
This is one possible way to amplify the stronger gaps in the magnitude-ordered eigenvalue spectrum. A large positive value of \eqref{eq:eigengap_detection} suggests then, as shown in Fig.~\ref{fig:eigenspectrum}, a number of coherent sets of $N_{CS}=n=82$ in Run 2, which is close to the estimate $( 2 \Gamma / \bar{\lambda}^{L})^{2} \approx 80$ (see Table I).

The key idea of SEBA \cite{Froyland2019} is to transform the eigenvectors $\{v_n\}_{n=1\dots N_{CS}}$  to a new set of vectors $\{z_k\}_{k=1\dots N_{CS}}$ which span the same subspace, but have significantly less non-vanishing components and thus disentangle the clusters of the graph better of each other. The $i$-th component of $z_k$ implies that network node ${\bm X}_i$ belongs to cluster $k$ with a likelihood $z_{k,i}$ when $z_{k,i}>0$ \cite{Froyland2019}. We combine this information in a vector ${\bm m}\in \mathbb{R}^N$ with $m_i = \max_{k=1,...,N_{CS}} z_{k,i}$ which results in an indicator of cluster affiliation (un)certainty. As shown below, the resulting soft classification allows for an improved cluster identification in comparison to the k-means method which was applied in our previous work \cite{Schneide2018}.

\begin{figure*}[t]
\begin{center}
\includegraphics[scale=1.0]{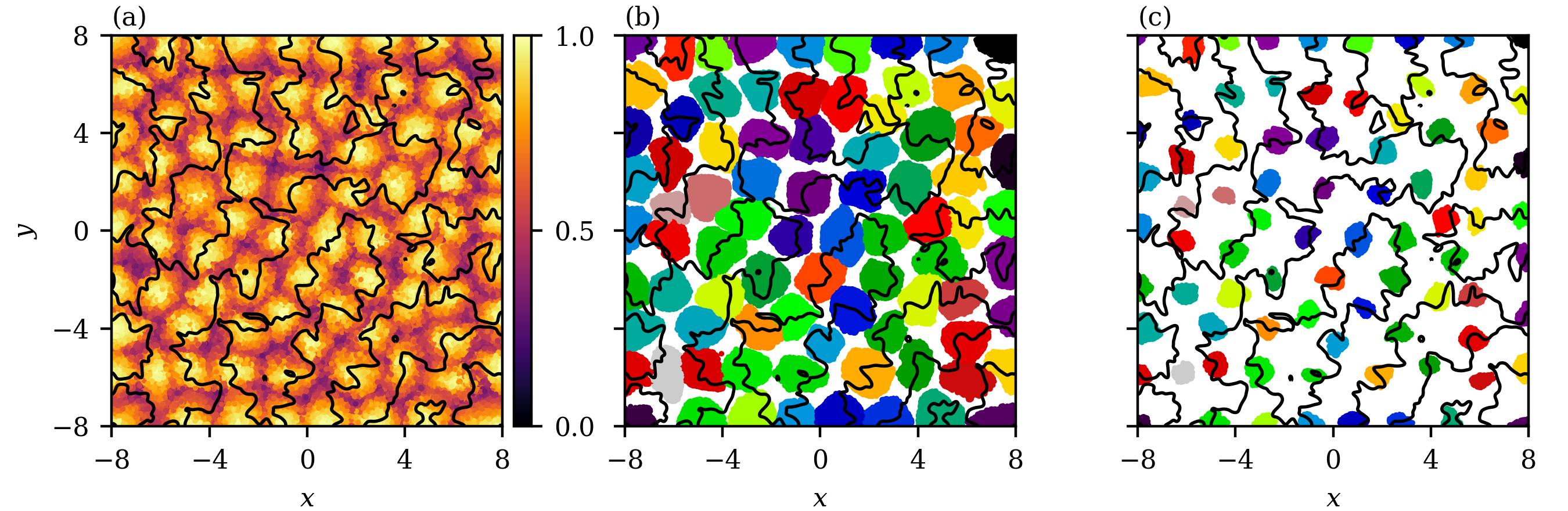}
\caption{Results of the sparse eigenbasis approximation (SEBA). (a) Scatter plot of maximum likelihood $m_i$ of tracer ${\bm X}_i (t_0+\Delta t/2)$ with $i=1,...,N$ to belong to one of the $N_{CS}=82$ clusters taken from the large value of $\chi_{82}$ in Fig.~\ref{fig:eigenspectrum}. Different clusters result for subsequent thresholding with $\zeta \approx 0.7$ in (b) and $\zeta=0.94$ in (c). Black lines indicate the isotherms $\langle T(z=0.5) \rangle_{\Delta t} = 0.5$.}
\label{fig:likelihood_clusters}
\end{center}
\end{figure*}

Figure \ref{fig:likelihood_clusters} displays the result of SEBA for the same data as in Fig.~\ref{fig:eigenspectrum}. Panel (a) shows a scatter plot of the maximum likelihood $m_i$ of tracer ${\bm X}_i$ with $i=1,...,N$ to belong to one of the $N_{CS}$ clusters. To separate the 82 features from each other, a threshold $\zeta$ has to be chosen to decompose the flow volume into features and an incoherent background. This idea of an incoherent background cluster was introduced in \cite{Hadjighasem2016}. Separating the CS properly requires $\zeta \approx 0.7$ and leads to panel (b) of Fig. \ref{fig:likelihood_clusters}. The CS intersect in most cases the isotherm (black solid line) of mean temperature $\langle T (z=0.5) \rangle_{\Delta t} = 0.5$, thus indicating that they represent the core regions of the large-scale circulation rolls which make up the turbulent superstructures in convection. When this threshold is raised to a higher value, such as $\zeta=0.94$ in panel (c), the clusters will be better separated. The threshold $\zeta$ is a free parameter, which allows us to control the separation of CS from the incoherent background, in contrast to the unsupervised $k$-means clustering for feature extraction from eigenvectors used in \cite{Hadjighasem2016,Banisch2017,Schneide2018,Schneide2019}. On the one hand, $\zeta$ should not be too small to separate the features properly. On the other hand, it should not be too close to 1 such that sufficiently many trajectory points can be assigned with a particular cluster.

\section{Analysis of Lagrangian heat transport}
\begin{figure}[t]
\begin{center}
\includegraphics[scale=1.0]{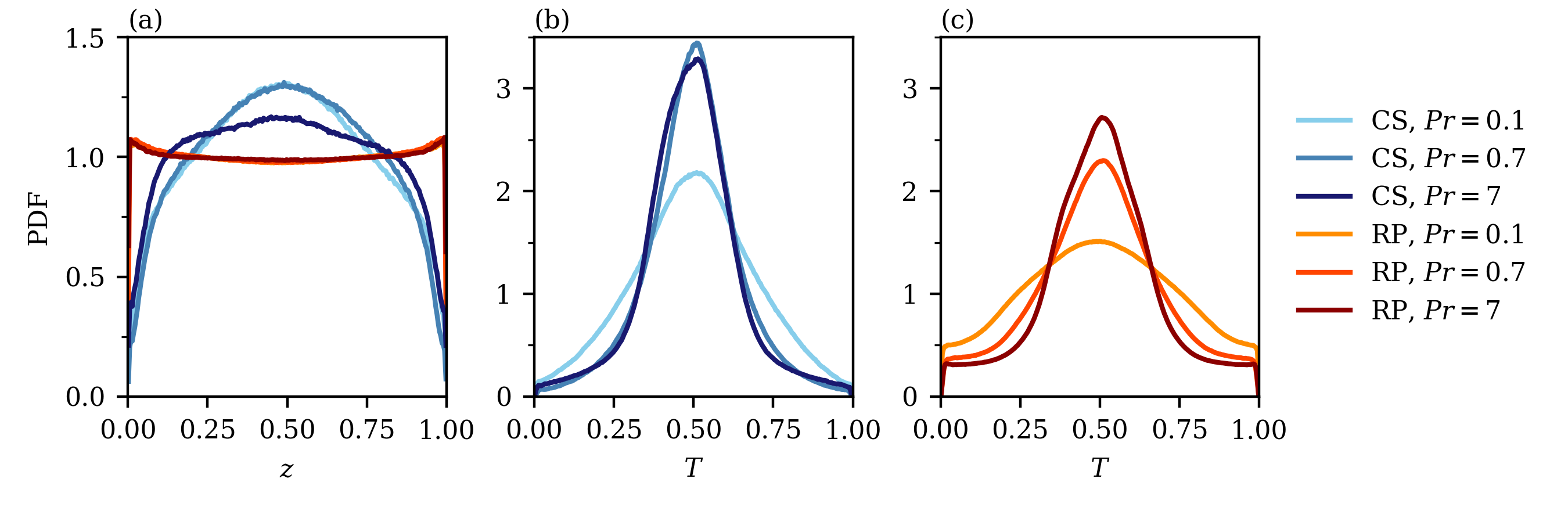}
\caption{Prandtl number dependence of the PDFs of tracer particles captured in the Lagrangian coherent sets (CS) and the spatial complement (RP). (a) PDF of the $z$-coordinate of the tracers in CS and RP. (b) PDF of the temperature $T$ along the trajectory in CS and (c) in RP. The PDFs are computed for $10$ disjoint time intervals of the evolution and arithmetically averaged subsequently. The threshold value is $\zeta=0.94$.}
\label{fig:PDFs_BL_separation}
\end{center}
\end{figure}

Figure \ref{fig:PDFs_BL_separation} confirms that the detected features (for $\zeta=0.94$) indeed represent the cores of convection rolls by providing the spatial distribution of the Lagrangian particles that belong to CS and comparing them with the remaining particles (RP) outside the identified sets. We display therefore the PDF of the vertical coordinate in panel (a) of the figure. The PDFs of the local temperature $T({\bm X}_i(t))$ that can be assigned with each Lagrangian tracer are shown in addition in panels (b,c). All three Prandtl numbers display clearly a narrower distribution around $T=0.5$ for the CS tracers, indicating that they are much less efficient in taking up and releasing heat as those outside the CS. The PDFs of $T$ are broadest for the smallest Prandtl number as the diffusion time is shortest \cite{Pandey2018}.

\begin{figure}[t]
\begin{center}
\includegraphics[scale=1.0]{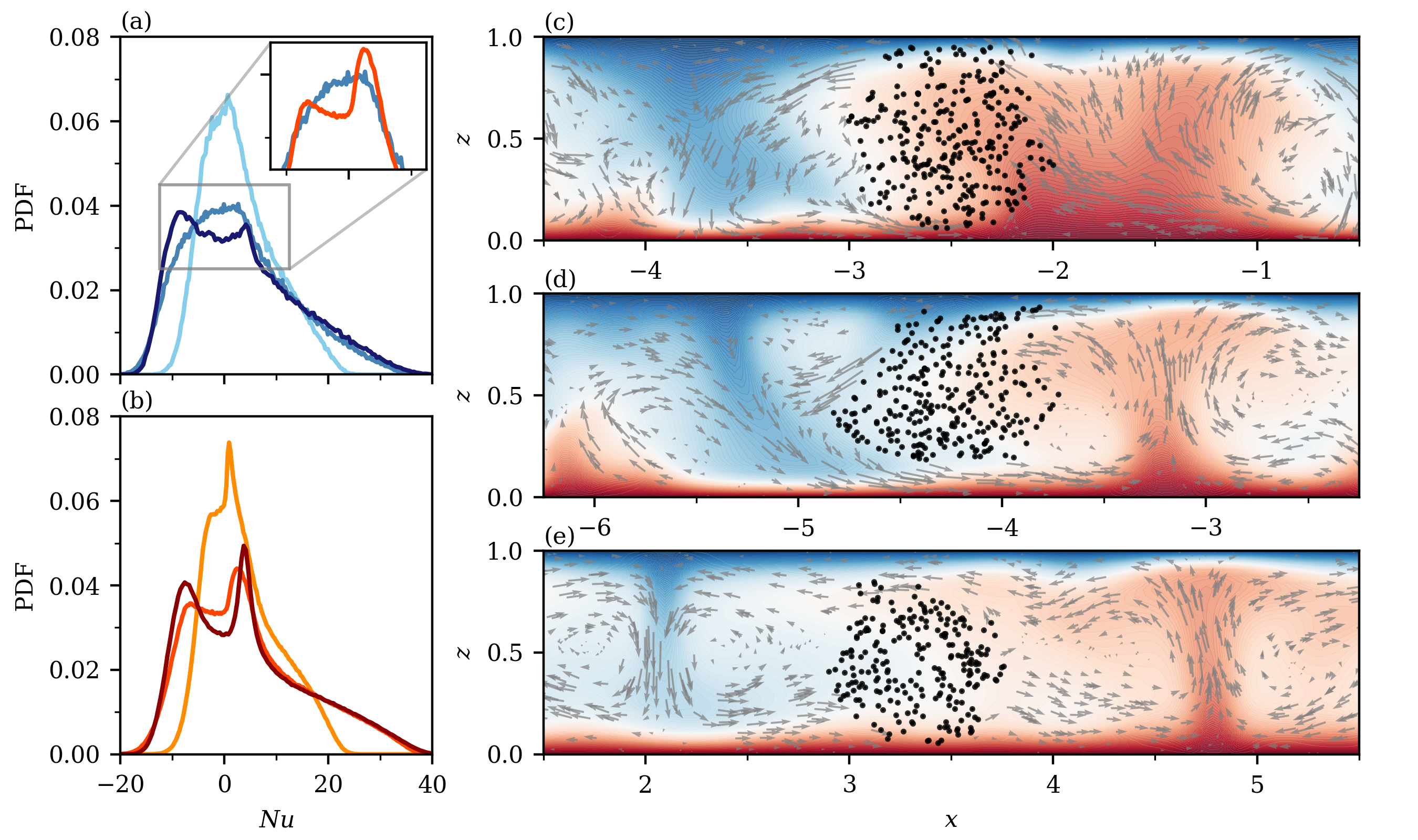}
\caption{Quantitative analysis and visualization of the contribution of Lagrangian CS to heat transfer. (a) PDF of $\langle \Nu_{\rm CS}^{L}\rangle_{\Delta t}$, (b) PDF of $\langle \Nu_{\rm RP}^{L}\rangle_{\Delta t}$. The inset in (a) is a one-to-one comparison of the central region of the PDFs for $\Pra = 0.7$. All colors equal that of Fig. \ref{fig:PDFs_BL_separation}. (c-e) Data in vertical cuts with a slice thickness of 0.2 through the convection layers are shown for $\Pra=0.1$ (c), $0.7$ (d), and $7$ (e). The corresponding temperature field, that is averaged over $\Delta t$, is shown as background. Tracers that belong to Lagrangian CS in this slice are indicated by black dots at time $t_0+\Delta t/2$. The grey vectors indicate velocity projections in the plane for tracers that belong to RP. The threshold value is again $\zeta=0.94$.}
\label{fig:PDFs_Nu}
\end{center}
\end{figure}

This suggests a closer look at the turbulent heat transport. While the global heat transfer is given by $\Nu^{E}=1+\sqrt{\Ra \Pr} \langle u_z T\rangle_{V,t}$ and listed in Table I, we can refine this dimensionless transport measure to disentangle the contributions of the Lagrangian CS and RP tracers. We therefore define a {\em local} Nusselt number which is proportional to the vertical component of the heat current vector along the Lagrangian trajectory ${\bm X}_i(t)$ \cite{Gasteuil2007,Schumacher2008,Schumacher2009} and given by
\begin{equation}
\label{eq:Lagrangian_Nu}
\Nu^{L}({\bm X}_i(t))  = \sqrt{\Ra\Pra}\, u_z T\Big |_{{\bm X}_i(t)} - \frac{\partial T}{\partial z}\Big |_{{\bm X}_i(t)} \,.
\end{equation}
Averaging over the time windows $\Delta t$ yields not only a Lagrangian perspective on $\Nu$, but as the individual particles represent specific spatial regions of the flow it also provides information on the Lagrangian CS and their spatial complement. This allows us to disentangle the heat transport into $\langle \Nu^{L}_{\rm CS}\rangle_{\Delta t}$ and $\langle \Nu^{L}_{\rm RP}\rangle_{\Delta t}$. The resulting PDFs for all three Prandtl numbers are displayed in Fig. \ref{fig:PDFs_Nu}. First it is observed in panels (a,b) that the support of the PDFs is smallest for the lowest Prandtl number confirming the reduced and less efficient heat transfer, which is in line with the coarsest thermal plumes shown in panel (c). The PDFs for the largest Prandtl number develop in both cases a pronounced bi-modal shape, in particular the PDF of $\langle \Nu^{L}_{\rm RP}\rangle_{\Delta t}$. While the peak at the positive local Nusselt number axis represents the strong plume detachment events from both boundary layers, the peak for the negative amplitude corresponds to impinging plumes. This effect is strongest for the largest $\Pra$ as thermal diffusion is smallest. While the differences between the distributions of $\langle \Nu^{L}_{\rm CS}\rangle_{\Delta t}$ and $\langle \Nu^{L}_{\rm RP}\rangle_{\Delta t}$ are highlighted for one $\Pra$ by the inset in panel (a), they become also prominent by the mean values of $\overline{\Nu}_{\rm CS}^L$ and $\overline{\Nu}_{\rm RP}^L$ in Table I for all $\Pra$. The latter is very close to the standard definition in the  Eulerian frame $\Nu^{E}=1+\langle u_z T\rangle_{V,t}$ \cite{Chilla2012}, but the former is smaller by about 1/3rd. Thus, our locally refined analysis of the turbulent heat transport demonstrates clearly that the cores of convection rolls, in which Lagrangian trajectories are trapped for longer times, contribute least to heat transport. One can unambiguously identify Lagrangian CS as spatial regions of reduced heat transport in the flow. This is furthermore underlined by comparing the mean fraction of tracers in CS, $\xi_{p}$, with their contribution to the global heat transport, $\xi_{q}$ (see Table I).

Panels (c-e) of Fig. \ref{fig:PDFs_Nu} visualize Lagrangian CS in instantaneous vertical cuts of the flows at all investigated $\Pra$. The Lagrangian CS are located between the rising and falling thermal plumes which develop thinner stems with increasing Prandtl number. These plume networks would be the structures that are identified as turbulent superstructures in the Eulerian frame of reference. The tracer cloud that forms the CS in these cases appears to be nearly elliptic and is thus characterized by a small surface-to-volume ratio.

\section{Summary and outlook}
We have identified Lagrangian coherent sets in three-dimensional turbulent Rayleigh-B\'{e}nard convection flows by means of clusters of tracer trajectories. A combination of the graph Laplacian framework, which originates from the time-averaged distances \cite{Hadjighasem2016} and uses a diffusion kernel with the subsequent sparse eigenbasis approximation  \cite{Froyland2019} led to an improved identification of the coherent sets in comparison to our previous analysis in \cite{Schneide2018}. It is shown that the majority of the clusters accumulates in the center of the layer between the spatial regions of strong up- or downwelling motions -- the latter of which would be identified as the significant feature in an Eulerian analysis of the flow.  We also showed that the Nusselt number $\overline{\Nu}_{CS}^L$, which can be assigned to CS in a Lagrangian way, is reduced to approximately 2/3rd of the standard Nusselt number $\Nu^{E}$ in all three cases. This quantifies that CS contribute significantly less to the global turbulent heat transfer compared to their spatial complement.

It is clear that these trajectories are not trapped for arbitrary long time intervals inside the Lagrangian coherent sets. Individual tracer particles will join the sets, which we illustrated in panels (c--e) in Fig. \ref{fig:PDFs_Nu}, others will leave them. Also an increase of the time interval $\Delta t$ will cause a shrinking of the set in space. A similar trend can be observed in \cite{Neamtu2019} for three-dimensional stably stratified flows or in \cite{Haller2020} for turbulent channel flows where Lagrangian coherent structures -- the two-dimensional manifolds that surround the Lagrangian coherent sets -- have been identified. For the present approach, this suggests to apply evolutionary cluster algorithms which would introduce a memory in time into the CS identification. These studies are currently underway and will be reported elsewhere. Network analyses have been recently applied in other turbulent flows to connect coherent structures and statistical flow properties in reduced models which provide a further direction to extend our present work \cite{Iacobello2021}.
    
\acknowledgments
The work of P.P.V. and C.S. is supported by the Deutsche Forschungsgemeinschaft with the Priority Programme DFG-SPP 1881 on Turbulent Superstructures. We thank Anna Kl\"unker, Gary Froyland and both reviewers for helpful comments. The authors gratefully acknowledge the Gauss Centre for Supercomputing e.V. (www.gauss-centre.eu) for funding this project by providing computing time through the John von Neumann Institute for Computing (NIC) on the GCS Supercomputer JUWELS at J\"ullich Supercomputing Centre (JSC).


\begin{thebibliography}{99}

\bibitem{Allshouse2015}
M. R. Allshouse and T. Peacock, 
Lagrangian based methods for coherent structure detection, 
Chaos {\bf 25}, 097617 (2015).

\bibitem{Haller2015}
G. Haller,
Lagrangian coherent structures,
Ann. Rev. Fluid Mech. {\bf 47}, 137 (2015).

\bibitem{Hadjighasem2017}
A. Hadjighasem, M. Farazmand, D. Blazevski, G. Froyland, and G. Haller,
A critical comparison of Lagrangian methods for coherent structure detection,
Chaos {\bf 27}, 053104 (2017).

\bibitem{FrLlSa10}
G. Froyland, S. Lloyd, and N. Santitissadeekorn,
Coherent sets for nonautonomous dynamical systems,
Physica D, 239 (2010).

\bibitem{Froyland2013}
G. Froyland, 
An analytic framework for identifying finite-time coherent sets in time-dependent dynamical systems, 
Physica D {\bf 250}, 1 (2013).

\bibitem{Karrasch2017}
D. Karrasch and J. Keller, 
A geometric heat-flow theory of Lagrangian coherent structures, 
J. Nonlinear Sci. {\bf 30}, 1849 (2020).

\bibitem{Froyland_Padberg_2015}
G.~Froyland and K.~Padberg-Gehle, 
A rough-and-ready cluster-based approach for extracting finite-time coherent sets from sparse and incomplete trajectory data, 
Chaos {\bf 25}, 087406 (2015).

\bibitem{Hadjighasem2016}
A. Hadjighasem, D. Karrasch, H. Teramoto, and G. Haller, 
Spectral-clustering approach to Lagrangian vortex detection, 
Phys. Rev. E {\bf 93}, 063107 (2016).

\bibitem{Banisch2017}
R. Banisch and P. Koltai,
Understanding the geometry of transport: Diffusion maps for Lagrangian trajectory data unravel coherent sets,
Chaos {\bf 27}, 035804 (2017).

\bibitem{Schlueter2017}
K. L. Schlueter-Kuck and J. O. Dabiri,
Coherent structure colouring: Ifigure5.pngdentification of coherent structures from sparse data using graph theory,
J. Fluid Mech. {\bf 811}, 468 (2017).

\bibitem{Padberg2017} 
K. Padberg-Gehle and C. Schneide, 
Network-based study of Lagrangian transport and mixing, 
Nonlin. Processes Geophys. {\bf 24}, 661 (2017).

 \bibitem{FroylandJunge2018}
G. Froyland and O. Junge, 
Robust FEM-based extraction of finite-time coherent sets using scattered, sparse, and incomplete trajectories, 
SIAM J. Appl. Dyn. Sys. (2018).

\bibitem{Schneide2018}
C. Schneide, A. Pandey, K. Padberg-Gehle, and J. Schumacher, 
Probing turbulent superstructures in Rayleigh-B\'enard convection by Lagrangian trajectory clusters,
Phys. Rev. Fluids {\bf 3}, 113501 (2018).

\bibitem{Schneide2019}
C. Schneide, M. Stahn, A. Pandey, O. Junge, P. Koltai, K. Padberg-Gehle, and J. Schumacher, 
Lagrangian coherent sets in turbulent Rayleigh-Bénard convection,
Phys. Rev. E {\bf 100}, 053103 (2019).

\bibitem{Kluenker2020}
A. Kl\"unker, C. Schneide, G. Froyland, J. Schumacher, and K. Padberg-Gehle,
Set-oriented finite-element study of coherent behaviour in Rayleigh-B\'{e}nard convection,
In: Junge O., Schütze O., Froyland G., Ober-Blöbaum S., Padberg-Gehle K. (eds) Advances in Dynamics, Optimization and Computation. SON 2020. Studies in Systems, Decision and Control, vol 304. Springer, Cham, pp. 86--108 (2020).

\bibitem{Ahlers2009} 
G. Ahlers, S. Grossmann, and D. Lohse,
Heat transfer and large scale dynamics in turbulent Rayleigh-B\'enard convection,
Rev. Mod. Phys. {\bf 81}, 503 (2009).

\bibitem{Chilla2012}
F. Chill\`{a} and J. Schumacher, 
New perspectives in turbulent Rayleigh-B\'{e}nard convection,
Eur. Phys. J. E {\bf 35}, 58 (2012).

\bibitem{Stevens2005}
B. Stevens, 
Atmospheric moist convection,
Annu. Rev. Earth Planet. Sci. {\bf 33}, 605 (2005).

\bibitem{Schumacher2020}
J. Schumacher and K. R. Sreenivasan,
Colloquium: Unusual dynamics of convection in the Sun,
Rev. Mod. Phys. {\bf 92}, 041001 (2020).

\bibitem{Kelley2018}
D. H. Kelley and T. Weier, 
Fluid mechanics of liquid metal batteries,
Appl. Mech. Rev. {\bf 70}, 020801 (2018).

\bibitem{Meila2001}
M. Meila and J. Shi,
A random walk view of spectral segmentation,
Proceedings of the Eighth International Workshop on Artificial Intelligence and Statistics, AISTATS 2001, Key West, Florida, USA, January 4-7, 2001, Eds. T. S. Richardson and T. S. Jaakkola, 6 pages (2001).

\bibitem{Coifman2005}
R. R. Coifman, S. Lafon, A. B. Lee, M. Maggioni, B. Nadler, F. Warner, and S. W. Zucker, 
Geometric diffusions as a tool for harmonic analysis and structure definition of data: Diffusion maps,
Proc. Natl. Acad. Sci. USA {\bf 102}, 7426 (2005).

\bibitem{Coifman2006}
R. R. Coifman and S. Lafon, 
Diffusion maps,
Appl. Comput. Harmon. Anal. {\bf 21}, 6 (2006).

\bibitem{Froyland2019}
G. Froyland, C. P. Rock, and K. Sakellariou, Sparse eigenbasis approximation: Multiple feature extraction across spatiotemporal scales with application to coherent set identification, Commun. Nonlinear Sci. Numer. Simulat. {\bf 77}, 81 (2019).

\bibitem{Fischer1997}
P. F. Fischer,
An overlapping Schwarz method for spectral element solution of the incompressible Navier-Stokes equations,
J. Comp. Phys. {\bf 133}, 84 (1997).
 
\bibitem{Scheel2013}
J. D. Scheel, M. S. Emran, and J. Schumacher, 
Resolving the fine-scale structure in turbulent Rayleigh-B\'{e}nard convection,
New J. Phys. {\bf 15}, 113063 (2013).

\bibitem{Stevens2018}
R. J. A. M. Stevens, A. Blass, X. Zhu, R. Verzicco, and D. Lohse, 
Turbulent thermal superstructures in Rayleigh-Bénard convection, 
Phys. Rev. Fluids {\bf 3}, 041501(R) (2018).

\bibitem{Pandey2018}
A. Pandey, J. D. Scheel, and J. Schumacher, 
Turbulent superstructures in Rayleigh-B\'{e}nard convection,
Nat. Commun. {\bf 9}, 2118 (2018).

\bibitem{Green2020}
G. Green, D. G. Vlaykov, J. P. Mellado, and M. Wilczek, 
Resolved energy budget of superstructures in Rayleigh–Bénard convection, 
J. Fluid Mech. {\bf 887}, A21 (2020).

\bibitem{Krug2020}
D. Krug, D. Lohse, and R. J. A. M. Stevens, 
Coherence of temperature and velocity superstructures in turbulent Rayleigh–Bénard flow, 
J. Fluid Mech. {\bf 887}, A2 (2020).

\bibitem{Vieweg2020}
P. P. Vieweg, J. D. Scheel, and J. Schumacher, Supergranule aggregation for constant heat flux-diven turbulent convection, Phys. Rev. Research {\bf 3}, 013231 (2021). 

\bibitem{Kluenker2019}
A. Kl\"unker, C. Schneide, A. Pandey, K. Padberg-Gehle, and J. Schumacher, 
Lagrangian perspectives on turbulent superstructures in Rayleigh‐Bénard convection, 
Proc. Appl. Math. Mech. {\bf 19}, 201900201 (2019).


\bibitem{Porte2008}
J. de la Porte, B. M. Herbst, W. Hereman, and S. J. van der Walt, 
An Introduction to Diffusion Maps, 
Proceedings of the Nineteenth Annual Symposium of the Pattern Recognition Association of South Africa, pp. 15--26 (2008).


\bibitem{Chi2007}
Y. Chi, X. Song, D. Zhou, K. Hino, and B. Tseng, 
Evolutionary spectral clustering by incorporating temporal smoothness, 
KDD '07: Proceedings of the 13th ACM SIGKDD international conference on Knowledge discovery and data mining, pp. 153--162 (2007).

\bibitem{Gasteuil2007}
Y. Gasteuil, W. L. Shew, M. Gibert, F. Chillá, B. Castaing, and J.-F. Pinton, 
Lagrangian temperature, velocity, and local heat flux measurement in Rayleigh-Bénard convection, 
Phys. Rev. Lett. {\bf 99}, 234302 (2007). 

\bibitem{Schumacher2008}
J. Schumacher, 
Lagrangian dispersion and heat transport in convective turbulence, 
Phys. Rev. Lett. {\bf 100}, 134502 (2008).

\bibitem{Schumacher2009}
J. Schumacher, 
Lagrangian studies in convective turbulence, 
Phys. Rev. E {\bf 79}, 056301 (2009).

\bibitem{Neamtu2019}
M. M. Neamtu-Halic, D. Krug, G. Haller, and M. Holzner, Lagrangian coherent structures and entrainment near the turbulent/non-turbulent interface of a gravity current,
J. Fluid Mech. {\bf 877}, 824 (2019).

\bibitem{Haller2020}
G. Haller, S. Katsanoulis, M. Holzner, B. Frohnapfel, and D. Gatti, Objective barriers to the transport of dynamically active vector fields, J. Fluid Mech. {\bf 905}, A17 (2020).

\bibitem{Iacobello2021}
G. Iacobello, L. Ridolfi, and S. Scarsoglio, A review on turbulent and vortical flow analyses via complex networks, Physica A {\bf 563}, 125476 (2021).
\end{thebibliography}
\end{document}